\documentclass[12pt]{article}
\usepackage{epsfig}
\usepackage{amsfonts}
\usepackage{latexsym}
\usepackage{amsmath}
\def\half{{1\over 2}}
\numberwithin{equation}{section}

\def\e{{\epsilon}}

\def\th{\theta}
 \def\p{\partial}

 \def\r{\rightarrow}
\def\hr{\hat{r}}

\def\s{\sigma}

\def\d{\delta}
\DeclareFontFamily{OT1}{rsfs}{}
\DeclareFontShape{OT1}{rsfs}{m}{n}{
<-7> rsfs5 <7-10> rsfs7 <10-> rsfs10}{}
\DeclareMathAlphabet{\mycal}{OT1}{rsfs}{m}{n}

\newcommand{\bea}{\begin{eqnarray}}
\newcommand{\eea}{\end{eqnarray}}
\newcommand{\be}{\begin{equation}}
\newcommand{\ee}{\end{equation}}
\newcommand{\non}{\nonumber}
\def\O{\mathcal{O}}

  \makeatletter
  \let\over=\@@over \let\overwithdelims=\@@overwithdelims
  \let\atop=\@@atop \let\atopwithdelims=\@@atopwithdelims
  \let\above=\@@above \let\abovewithdelims=\@@abovewithdelims
  \makeatother
\begin{document}

\begin{titlepage}
\vskip 3 cm
\begin{center}

{}

 {\Large \bf Microscopic Realization of the Kerr/CFT Correspondence}

\vskip 1.5cm {Monica Guica $^\S$ and Andrew Strominger$^\dag$}
 \vskip 1.5 cm

{\it $^\dag$Center for the Fundamental Laws of Nature\\
    Jefferson Physical Laboratory, Harvard University, Cambridge, MA 02138, USA}\\

\vspace{0.5cm}

{\it $^\S$ Institut de Physique Th\'eorique \\
   CEA Saclay, CNRS-URA 2306,
  91191 Gif sur Yvette, France
} \\

\end{center}

\vskip 0.6cm

\begin{abstract}

\vskip 0.5cm

Supersymmetric M/string compactifications to five dimensions contain BPS black string solutions with magnetic 
graviphoton charge $P$ and near-horizon geometries which are quotients of $AdS_3\times S^2$. The holographic duals 
are typically known 2D CFTs with central charges $c_L=c_R=6P^3$ for large $P$. These same 5D compactifications also 
contain non-BPS but extreme Kerr-Newman black hole solutions with
$SU(2)_L$ spin $J_L$ and electric graviphoton charge $Q$ obeying $Q^3 \le J_L^2$. It is shown that in the maximally 
charged limit $Q^3\to J_L^2$, the near-horizon geometry
coincides precisely with the right-moving temperature $T_R=0$ limit of the black string with magnetic charge
 $P=J_L^{1/3}$. The known dual of the latter is identified as  the $c_L=c_R=6J_L$ CFT predicted by 
the Kerr/CFT correspondence. Moreover,
at linear order away from maximality, one finds a $T_R \neq 0$ quotient of the $AdS_3$ factor of the black string
 solution and the associated thermal CFT entropy reproduces the linearly sub-maximal Kerr-Newman entropy. Beyond linear order,  for general $Q^3<J_L^2$,
one has a finite-temperature quotient of a warped deformation of the magnetic string geometry. The corresponding dual deformation of the magnetic string CFT potentially supplies, for the general case, the $c_L=c_R=6J_L$ CFT predicted by Kerr/CFT.


\noindent

\end{abstract}

\vspace{3.0cm}

\end{titlepage}
\pagestyle{plain}
\setcounter{page}{1}
\newcounter{bean}
\baselineskip18pt


\setcounter{tocdepth}{2}
\tableofcontents

\section{Introduction}

A complete accounting of the quantum microstates of a black hole requires a complete quantum theory
 of gravity. Such a complete accounting was obtained for certain supersymmetric black holes 
\cite{ascv} by using string theory to identify their microstates with those of a dual two-dimensional conformal field 
theory (2D CFT).  Later it was realized \cite{Strominger:1997eq}
that notable features of the construction - in particular, matching the universality of the 2D CFT Cardy formula to the
 universality of the Bekenstein-Hawking area law - followed largely from a careful analysis \cite{Brown} of the 
properties of the diffeomorphism group in the $AdS_3$ region near the horizon.  This implies that the leading order
agreement of the microscopic and macroscopic pictures does not require any fine details of string theory. Rather, 
it must follow a similar pattern for black holes with $AdS_3$ near-horizon regions in any consistent ultraviolet 
completion of quantum gravity.\footnote{Of course it is a logical possibility that string theory is the only such 
completion.} This is of course in accord with the expectation that universal long-distance phenomena should not depend 
on the details of 
short-distance microphysics. It has nevertheless been essential to our understanding to have the specific microscopic
 examples in string theory \cite{ascv} in which the general arguments of \cite{Strominger:1997eq} are given a 
concrete and fully self-consistent realization.

Recently, a general analysis of the near-horizon diffeomorphisms  of astrophysical Kerr black 
holes with near-maximal spins, patterned after \cite{Brown,Strominger:1997eq}, has similarly 
indicated that they are  dual to 2D CFTs \cite{Guica:2008mu}, but so far there has been no 
proposed string theoretic realization of the duality. Evidence for the conjectured Kerr/CFT 
correspondence is given by the agreement of the Cardy formula with the area law,  the CFT 
correlators with near-superradiant  Press-Teukolsky greybody factors 
\cite{Teukolsky:1973ha,
Press:1973zz,Teukolsky:1974yv, Bredberg:2009pv}, and many other analyses
 \cite{Hartman:2009nz} -
\cite{Dias:2009ex}. However, there are several puzzling features\footnote{For example 
the appearance of complex conformal weights or the apparent lack of boundary conditions admitting
 two Virasoros.} which are hard to understand without a concrete microscopic construction. It is 
the purpose of this paper to fill this gap by proposing such a construction for the general extreme
 five-dimensional Kerr-Newman black hole embedded as a solution of string theory.

The basic idea is as follows. Supersymmetric M/string compactifications to five dimensions
have BPS string solutions with magnetic graviphoton charge $P$ and a near-horizon
$AdS_3\times S^2$ geometry. Typically, the dual CFT is  explicitly known from a brane construction and has central charges
$c_L=c_R=6P^3$ for large $P$. We will refer to it as the ``$P^3$ CFT".  For the special case of M-theory on a Calabi-Yau,
the $P^3$ CFT is the worldvolume theory of an M5-brane wrapping the 4-cycle associated to $P$, and is often referred to as the MSW CFT \cite{msw}.  We shall be primarily interested in the case where the Calabi-Yau contains a torus factor and there is an uplift from five to six dimensions using the graviphoton fiber.  The total space of the graviphoton fiber over the near-horizon
$AdS_3\times S^2$ geometry is then a quotient of
$AdS_3\times S^3$ \cite{spanishdudes, mgas}.

The same compactifications also contain non-supersymmetric but extreme Kerr-Newman black
holes with $SU(2)_L$ spin $J_L$ and electric graviphoton charge obeying $Q^3\le J_L^2$.
\footnote{$Q=0$ is just  5D extreme Kerr, and we shall refer to the $Q^3=J_L^2$ limit as the
 maximal case.}
These all have a near-horizon
$AdS_2\times S^3$ geometry where the $S^3$ is fibered over the $AdS_2$ and squashed, known as the 5D NHEK-Newman geometry. Including the graviphoton fiber
we get a warped/squashed
$AdS_3\times S^3$ geometry.

The key observation is that in the maximally charged limit $Q^3\to J_L^2$, the squashing and warping
go away and the near-horizon NHEK-Newman geometry coincides exactly with  the
near-horizon geometry of the charge $P=J_L^{1/3}$ magnetic string\footnote{In order for this to 
make sense, $J_L$ must be the cube of an integer. However this is not a restriction in the 
semiclassical limit of interest here because,  restoring the factors of $\hbar$, the spacing
 $\Delta J_L$ between two allowed values of $J_L$ goes as
 $\Delta J_L \sim \hbar^{1/3}J_L^{2/3}$. Hence the allowed values are effectively continuous for $\hbar \to 0$.}, 
with a quotient which corresponds to a
right-moving temperature $T_R=0$ on the string.  Hence the maximal Kerr-Newman black hole is also dual to the $P^3$ CFT!\footnote{A very similar relation to the supersymmetric $J_L^2<Q^3$ BMPV black holes for $J_L^2\to Q^3$ was found in \cite{mgas}.} The values $c_L=c_R=6P^3=6J_L$ for this CFT agree with the predictions of  the Kerr/CFT correspondence.  The full asymptotically flat black hole and black string geometries are of course very different.  The gluing of the near-horizon region into the asymptotically flat region of the black hole geometry 
 differs from that of the black string geometry  by an exchange of the fiber and spatial angular coordinate.  This results in  very different dictionaries relating the spacetime and  $P^3$ CFT quantum numbers.

Once we have identified the holographically dual CFT for maximal Kerr-Newman, the duals for the sub-maximal extreme 
Kerr-Newman can be found by matching the deformation of the bulk geometry to one of the boundary CFT.
At linear order, the only effect relevant for the entropy computation is that the action of the quotient is changed
 in a manner that corresponds to a
non-zero right-moving temperature $T_R$. Inserting this value of $T_R$ into  Cardy's formula perfectly reproduces the 
area law, in accord with the conjectured Kerr/CFT correspondence. At second order, a new effect is relevant:  
the $AdS_3$ factor becomes warped due to a non-zero ``self-dual" electric graviphoton field which preserves 
$SL(2,R)_L$ but breaks $SL(2,R)_R$ to $U(1)$. In principle this should be describable directly as some kind of 
continuous deformation of the dual $P^3$ CFT and quantum state, but it is beyond the aspirations of the present 
paper to do so.  The Kerr/CFT analysis indicates that the deformed theory exists, and is a CFT whose Cardy entropy 
matches the area law.   However, now that we have embedded Kerr/CFT duality  into string theory, standard string theory techniques should be applicable to work out the detailed form of the holographic dual and microscopically verify the Kerr/CFT  picture.  We will return to the central problem of understanding  the deformations beyond linear order away from maximality from the CFT side elsewhere \cite{ghss}.

An important feature of the finitely-deformed geometry is that it is no longer possible (at least in the most naive fashion) to glue on an asymptotically flat region suitable for a 5D black string.  Gluing on the asymptotically flat Kerr-Newman region requires a specific algebraic relation correlating the temperature $T_R$ with the warping factors. This gluing constraint is essential for the consistency of the whole picture: otherwise  the number of CFT and black hole  parameters would fail to match.

This paper is organized as follows. In section 2 we state the conjecture. In section 3 we 
review the geometry and all necessary formulae for the extreme 5D Kerr-Newman solution, as well as its 6D lift. 
Section 4 reviews the Kerr/CFT analysis which leads to an asymptotic symmetry group gnerated by a  $c=6J_L$ Virasoro. 
 In section 5, it is shown that for maximal spin the bulk geometry locally acquires an AdS$_3$ factor and we can 
thereby identify the boundary $c=6J_L$ CFT dual. An explicit example of a brane construction in which the microscopic dual is the MSW string arising from 3 intersecting stacks of P M5 branes in a $T^6$ compactification of M theory is worked out in detail. 
Linear deformations from maximality  are analyzed, agreement between the bulk and boundary properties is demonstrated and the general non-maximal case is discussed. 

Some comments on the existence of different dual CFTs related to the  long and short string pictures are made in an appendix.

\section{The conjecture}

Five-dimensional supergravity has black hole solutions characterized by a mass $M$, half-integral $SU(2)_L\times SU(2)_R$ spins
 $J_L$ and $J_R$
and graviphoton charge $Q$. In this paper we consider a two-parameter family of non-supersymmetric extremal spinning 
black holes characterized by

\be
J_R =0 \;,\;\;\;\;\; Q^3 \leq J_L^2 \;,\;\;\;\;\;   \left(\frac{M}{3} - \frac{Q}{2} \right)^2
\left(\frac{2M}{3}+ 2Q\right) =J_L^2. \label{ppo}
\ee
The Bekenstein-Hawking area law yields the entropy

\be  S_{BH}=2\pi\sqrt{J_L^2-Q^3}.\ee These black holes are distinct from the supersymmetric BMPV black holes which can have the same spin but must have $Q^3\ge J_L^2 $ and $M=3Q$.

The theory also contains supersymmetric black string solutions characterized by magnetic graviphoton charge $P$ and a near-horizon $AdS_3\times S^2$ geometry.  When this is embedded in M/string theory the dual 2D ``$P^3$" CFT is known in many cases \cite{msw} and
has left and right central charges for large $P$
\be c_L=c_R=6 P^3 .
\ee
The conjecture is that the extremal black holes obeying (\ref{ppo}) with spin
\be
\label{llp} J_L=P^3,
\ee
are dual to a deformation of the known $P^3$ CFT at temperatures
\be
\label{trconj}
T_L=0,~~~~~T_R={\sqrt{1-{Q^3 \over J_L^2}} \over \pi}.
\ee
The CFT is deformed by the addition of certain operators to the action, described 
herein in the bulk picture in terms of a continuous warping of the near-horizon $AdS_3\times S^2$ geometry. Note that $c_R=c_L=6J_L$ as expected in the Kerr/CFT correspondence.

Our conjecture is best-motivated in the extended supersymmetric cases of M-theory on $T^6$ or $K3\times T^2$. However, one may consider a more general compactification of M theory on a Calabi-Yau $X$, which give rise 
to $\mathcal{N}=2$ 5d supergravity coupled to $n-1$ vector multiplets. The vector moduli space of the compactification 
is parametrised by scalars $t^A$, $A=1, \ldots n$, which satisfy
\be
D_{ABC}\, t^A t^B t^C =1
\ee
where $6 \,D_{ABC}$ are the triple self-intersection numbers of the four-cycles of $X$. The black 
hole is charaterised by $n$ electric charges $q_A$ and the angular momentum $J_L$. At the horzion, the moduli
 must take attractor values which obey
\be
q_A = 3\, Q D_{ABC}\, t^B t^C
\ee
where $Q$ is the graviphoton charge. For this setup, our proposal is that the black hole is described by a
deformation of the MSW CFT associated to wrapping an M5-brane on the cycle
\be
 p^A = J_L^{1/3} t^A \label{zzp} \;, \;\;\;\;\; D_{ABC}\, p^A p^B p^C = J_L
 \ee
This MSW CFT has central charges
\be
c_L=c_R= 6\, D_{ABC}\,p^Ap^Bp^C=6\,J_L
\ee
plus corrections which are subleading for large $J_L$. The extremal black hole is then conjecturally dual to a thermal state in the deformed CFT in which right-movers are excited to the temperature $T_R$ given in \eqref{trconj}.

Note that the M5-brane wraps the four-cycle whose self-intersection is the two-cycle associated to the M2 charge carried by the black hole in the maximal case $Q^3=J_L^2$, and hence has the same attractor point (\ref{zzp}). 
 Of course not every two-cycle is the self intersection of a four-cycle: our conjecture in its present form pertains 
only to such cases.


\section{Geometry}

Five dimensional supergravity contains the universal
gravity-graviphoton sector

\be \label{fd} S_5={1\over 4\pi^2}\int d^5 x \left(\sqrt{-g}(R- \frac{3}{4}F^2)+{1\over 4}
 \epsilon^{abcde}A_aF_{bc}F_{de}\right).\ee
We choose conventions with $G_5={\pi \over 4}$, $\e^{tr\th\psi\phi}=1$ and 
 ${1 \over 2\pi}\int F \in \mathbb{Z}$.  Dirac quantization then implies the electric 
charge $Q$ defined here as  $Q\equiv {1 \over 4\pi^2}\int *F$ obeys $3 Q \in \mathbb{Z}$.

\subsection{Full solution}
We consider a five-dimensional charged rotating black hole solution parameterized by the mass M, $SU(2)_L$ spin $J_L$ and 
graviphoton charge $Q$. These can be traded for another set of three parameters $a, M_0$ and $\delta$ via the relations

\bea
M &=& {3M_0\over 2} \cosh 2\delta \non
\\
J_L  &=& a M_0\,(c^3 + s^3 ),\non
\\
Q  &=& M_0sc
\eea
where $c = \cosh \delta$ and $s = \sinh \delta$. The black hole solution reads
\cite{Breckenridge:1996is,Cvetic:1996xz,Dias:2007nj}

\bea
\label{5dmet}
ds_5^2 & = & - \frac{(a^2+\hr^2)(a^2+\hr^2-M_0) }{\Sigma^2}\, d\hat{t}^2 + 
\Sigma \left(\frac{\hr^2 d\hr^2}{f^2-M_0 \hr^2}+ \frac{d\th^2}{4}\right) - \frac{M_0 F}{\Sigma^2} \,
 (d\hat{\psi}+\cos\th\, d\hat{\phi})\, d\hat{t} \non \\
&& + \frac{\Sigma}{4} (d\hat{\psi}^2+d\hat{\phi}^2+2\cos\th \, d\hat{\psi}\, d\hat{\phi}) 
 + \frac{a^2 M_0 B}{4\Sigma^2}\, (d\hat{\psi}+\cos\th \,d\hat{\phi})^2
\eea
\be
 A= \frac{M_0  \sinh 2\delta }{2 \Sigma} \left( d\hat{t} - \half a e^\delta (d\hat{\psi} + \cos\th \, d\hat{\phi}) \right)
\ee
and we defined the quantities
\be
B= a^2+\hr^2 -2 M_0 s^3 c^3 - M_0 s^4(2 s^2+3)\;, \;\;\;\;\;\; F= a(\hr^2+a^2) (c^3+s^3) - a M_0 s^3
\ee
\be
\Sigma =  \hat{r}^2 + a^2 + M_0 s^2  \ , \quad f = \hat{r}^2 + a^2 \ ,
\ee 
We also note that the $SU(2)_L$ angle $\hat \psi$ is identified as
\be \hat \psi \sim \hat \psi+4\pi n.\ee
There is a more general solution with two independent angular momenta, but we have set the other angular momentum, $J_R$, to zero. For this case, the  $SU(2)_R$ rotational symmetry considerably simplifies matters.

The surface gravities at the inner and outer horizons $r_\pm^2 = \frac{1}{2} (M_0-2a^2) \pm \frac{1}{2} \sqrt{M_0(M_0 - 4a^2)}$ are
\be
\frac{1}{\kappa_{\pm}} = \frac{\sqrt{M_0}}{2} \left((c^3+ s^3) \pm \frac{c^3- s^3}{\sqrt{1-4 a^2 /M_0}}  \right).
\ee
in terms of which the Hawking temperature is
$ T_H={\kappa_+ \over 2 \pi}.$ The horizon angular velocities are
\be
\Omega_L = \Omega_\phi - \Omega_\psi = \frac{4a}{M_0}\left[ (c^3- s^3)+(c^3+s^3) \sqrt{1- 4 a^2 / M_0}\right]^{-1} \;, \;\;\;\;\; \Omega_R=0,
\ee
and the electric potential at the horizon  reads
\be
\Phi = \frac{c^2 s -s^2 c + (c^2s+s^2c) \sqrt{1- 4 a^2 / M_0}}{c^3-s^3 + (c^3+s^3)\sqrt{1- 4 a^2 / M_0} }.
\ee

\subsection{Six-dimensional lift} 
Any solution of the 5D bosonic minimal supergravity action (\ref{fd}) can be lifted 
(see \cite{Duff:1998cr} for details) to a solution of the 6D theory 
\be
 \label{sd} S_6={ 1 \over 8\pi^3}\int d^6 x\sqrt{-g}\left(R- \frac{1}{12}H^2\right),
 \ee
where the three-form field strength $H$ is restricted by the self-duality conditions 
$H=*H$.  The lift is given by 
\bea 
\label{6dlift} ds_6^2&=&ds_5^2+(d\hat u+A)^2, \\
         H&=&-dA\wedge(d\hat u+A)-*\bigl(dA\wedge(d\hat u+A)\bigr),
         \eea
         where 
\be
 \label{id} \hat u \sim \hat u +2\pi m
\ee 
will be referred to as the $U(1)$ fiber coordinate.
The lift of the Kerr-Newman geometry is
\bea
\label{bh5metric}
ds_6^2  &=& -\left(f - M_0 c^2 \right){d\hat{t}^2\over \Sigma} +{d\hat u^2}  +M_0 {\sinh 2\delta\over \Sigma}\,
d\hat{t} d\hat u +
 \Sigma\left( {\hat{r}^2 d\hat{r}^2\over f^2 - M_0 \hat{r}^2} + \frac{d\theta^2}{4}\right)\notag\\
& & + \frac{\Sigma}{4}(\,d\hat{\psi}^2 +  d\hat{\phi}^2 + 2 \cos\theta \,d\hat{\phi} \,d \hat \psi)
 + {M_0 a^2\over 4 \Sigma}(d\hat{\psi} + \cos\theta \,d\hat{\phi})^2 \notag \\
& & - {M_0 a\over \Sigma}\left( (c^3+ s^3)\,d\hat{t} + (s^2 c + c^2 s)\,d\hat u\right)( d\hat{\psi} + \cos \theta \,d\hat{\phi})
\eea
For many questions, the 6D picture is simpler than the 5D one, and we shall  specialize to this case when convenient.
 In particular, the interchange of fiber and angular coordinates which relates black string and black hole descriptions 
is simply a diffeomorphism when there is a 6D lift, whereas in the more general context an M/string duality may be 
needed to justify the interchange.

\subsection{Extremality}

We are interested in
a family of extremal, non-supersymmetric black holes obtained by setting
\be
M_0 = 4a^2 \ .
\ee
In this case
\bea
M &=& {6a^2} \cosh 2\delta \non
\\
J_L  &=&  4a^3(c^3+ s^3),\non
\\
Q  &=&4a^2 s c  \ ,
\eea
the entropy is
\be S_{BH}=8\pi a^3(c^3-s^3)=2\pi \sqrt{J_L^2-Q^3} \label{sbzh} \ee
and
\bea
\Phi &=& {c^2 s-s^2c \over
c^3-s^3} \\
\Omega_L &=& {1\over  a(c^3- s^3)}  \ .
\eea
According to the usual thermodynamic arguments, the quantum state  is mixed and
 given by the density matrix with eigenvalues
\be
 e^{-\beta_H(\omega-m\Omega_L+q\Phi)} 
\ee
where $\omega$, $m$ and $q$ are the energy, $J_L$ spin and charge of the state. Determining the density matrix at 
extremality is slightly subtle because $\beta_H\to \infty$. This implies only states with
 $\omega=m\Omega_L(M_0=4a^2)-q \Phi(M_0=4a^2)$ can contribute.  The spin and charge  dependence of the remaining 
density matrix on this subspace is then\footnote{We have chosen $T_R$ to be the variable conjugate to $2m$, rather 
than $m$, because for neutral Kerr black holes $m\sim - i \p_{\hat \psi}$ takes half integer values 
and $\hat \psi$ runs from $0$ to $4\pi$.  Hence it is $2m$ which will eventually become, in section 5,  the 
canonically normalized $L_0$ eigenvalue of a 2D CFT.  Similarly there is a factor of 3 in the definition of $T_Q$ 
because $3q \sim - i\p_{\hat u}$ is an integer. }
\be 
e^{-{2m \over T_R}-{3 q\over T_Q}},\ee
where
\bea 
{\label{tr}T_R}&=&-2\left[\beta_H (\Omega_L(M_0)-\Omega_L(4a^2) ) \right]^{-1}_{M_0=4a^2}={c^3-s^3 \over \pi(c^3+s^3)}={\sqrt{1-{Q^3 \over J_L^2}} \over \pi},
\\ {T_Q}&=&3\left[\beta_H (\Phi(M_0)-\Phi(4a^2) ) \right]^{-1}_{M_0=4a^2}={c^3-s^3 \over 
4\pi a s^2c^2}={\sqrt{J_L^2-{Q^3 }} \over  \pi Q^2}={J_LT_R \over  Q^2}. \eea
The reason for the notation ``$T_R$'' for the temperature conjugate to angular momentum will become apparent shortly.

\subsection{Near-horizon limit}

In this subsection we describe the near-horizon limit, following \cite{Bardeen:1999px,Dias:2007nj,Chen:2009xja}
To reach the near-horizon limit from (\ref{bh5metric}), define
\begin{equation}
t = {\Omega_L\over 2a^2} \hat{t} \epsilon \ , \quad \quad r = {\hat{r}^2 - a^2\over \epsilon} \ ,
\quad \quad y = {2 \pi T_Q}(\hat u + \Phi \hat{t}) \ , \non
\ee

\be
\psi = \hat{\psi} - \Omega_L \hat{t}-{2J_L\over Q^2}(\hat u + \Phi \hat{t}) \ , \quad \quad
\phi = \hat{\phi}  \label{nhlim6d}
\end{equation}
The near-horizon metric is
\bea\label{ff}
{12\over M}ds^2  &=& -r^2 dt^2 + {dr^2\over r^2} + \gamma(dy+rdt)^2 + \gamma(d\psi + \cos\theta d\phi)^2 \\
& & + 2 \alpha \gamma (dy + r dt)(d \psi + \cos\theta d\phi) + d\theta^2 + \sin^2\theta d\phi^2\notag
\eea
where the deformation parameters
\bea\label{squashparam}
\alpha &=& {2 \cosh 2 \delta \over 1+ \cosh^2 2 \delta}\notag\\
\gamma &=& 1 + {1\over \cosh^2 2 \delta} \ .\notag
\eea
are related by $M\alpha\gamma=12a^2$.

In terms of the $SL(2,R)_L \times SU(2)_R$ invariant forms,
\bea
\sigma_1 &=& \cos\psi d\theta + \sin\theta \sin\psi d\phi\\
\sigma_2 &=& -\sin\psi d\theta + \sin\theta \cos\psi d\phi\notag\\
\sigma_3 &=& d\psi + \cos\theta d\phi\notag\\
w_{\pm} &=& -e^{\mp y}rdt \mp e^{\mp y}dr/r\notag\\
w_3 &=& dy + r dt \ ,\notag
\eea
the metric can be written in the manifestly  $SL(2,R)_L \times SU(2)_R$ -invariant form
\bea\label{nicemetric}
{12\over M}ds^2 &=& -w_+ w_- + \gamma w_3^2+ \sigma_1^2 + \sigma_2^2  + \gamma \sigma_3^2 +2 \alpha \gamma w_3 \sigma_3\ .
\eea
 The gauge field $A$ is 
conveniently expressed via
\be 
d\hat u+A= -{a \over 2}\tanh 2 \delta \bigl(e^\delta\sigma_3+e^{-\delta}w_3\bigr)
\ee
and $H$ is then readily constructed from this expression via (\ref{6dlift}) as 
\begin{equation} \label{hf}
H=  \frac{Q}{4}  \left( \s_1 \wedge \s_2 \wedge \s_3 + \frac{1}{2} \, w_+ 
\wedge w_-\wedge w_3  + {\rm sech} 2 \delta \, ( \s_1 \wedge \s_2 \wedge w_3 + 
\frac{1}{2} \, w_+ \wedge w_-\wedge 
\s_3 )\right)
\end{equation}
The $(y,\psi )$
identifications are
    \be \label{id} y \sim y+{4 \pi^2T_Q m },~~~~~\psi \sim \psi-{4 \pi J_L m\over Q^2}+4\pi n, \ee for any integers $(m,n)$. The $U(1)$ fiber lies along the vector field
\be \label{fdir} f\equiv \p_{\hat u}={2 \pi T_Q }\,\p_y-{2J_L\over Q^2}\p_\psi,\ee
while gauge-invariant rotations along $J_L$ are generated by  
\be\label{jdir}j_L\equiv \p_{\hat \psi}-A_{\hat \psi}f=   \frac{2(c^3-s^3)}{\sinh 4\d} \left(
e^\d \p_y - e^{-\d}\p_\psi  \right),\ee
where we have used $A_{\hat \psi}=A_\psi=-{1 \over 2} \, a \,e^\delta \tanh 2\d$.

\section{The asymptotic symmetry group}

In this section we briefly review the computation of  \cite{Guica:2008mu, Compere:2009dp,Azeyanagi:2008kb,Hartman:2008pb}
 showing that the asymptotic symmetry group of the near-horizon spacetime \eqref{ff} with boundary conditions which surpress deviations from extremality consists of a centrally-extended Virasoro algebra with central charge
$ c_R = 6 J_L. $
In the convention of this paper, the chiral half of the CFT excited in the case of extreme black 
holes consists of right-movers.  This convention was chosen because this is usually the 
supersymmetric side and the excitations we consider break supersymmetry.

The near-horizon metric of the 5d black hole is
\be
ds_5^2 = \frac{M}{12} \left[-r^2 dt^2 + \frac{dr^2}{r^2} +d\theta^2 +\sin^2\theta d\phi^2 +\frac{27 J_L^2}{M^3} \left(d\tilde y+  \pi T_R \cos \th\, d\phi +rdt \right)^2\right]
\ee
where $\tilde y=\pi T_R (\hat \psi-\Omega_L\hat t)=y+\pi T_R \psi$ is identified as \be \label{diu}\tilde{y} \sim \tilde{y} + 4 \pi^2 T_R n .\ee  The near-horizon gauge field reads
\be \label{aeq}
A=-\half a e^\d \tanh 2 \d \left( \frac{d\tilde{y}}{\pi T_R} + \cos\th d\phi + e^{-2 \d} r dt \right)
\ee
 The boundary conditions on the deviations $\d g_{\mu\nu}$ from the background are  \cite{Compere:2009dp} 
\bea \label{bc}
&&\d g_{tt} \sim \O(r^2) \;, \;\;\;\;\d g_{t r} \sim \O(r^{-2})\;, \;\;\;\;\; \d g_{t \theta} \sim \O(r^{-1})\;, \;\;\;\;\; \d g_{t\tilde y } \sim \O(1) \;, \;\;\;\;\; \d g_{t\phi} \sim \O(r) \non \\
&& \d g_{rr} \sim \O(r^{-3})\;, \;\;\;\;\; \d g_{r\th} \sim \O(r^{-2}) \;, \;\;\;\;\; \d g_{r\tilde y } \sim \O(r^{-1}) \;, \;\;\;\;\; \d g_{r\phi} \sim \O(r^{-2}) \non \\
&&\d g_{\th \th} \sim \d g_{\th \tilde y } \sim \d g_{\th \phi} \sim \O(r^{-1}) \;, \;\;\;\;\; \d g_{\tilde y \tilde y } \sim \d g_{\tilde y  \phi} \sim \O(1) \;, \;\;\;\;\; \d g_{\phi\phi} \sim \O(r^{-1})
\eea
while boundary conditions on the gauge field read\footnote{The boundary condition on 
$A_{\tilde y}$  can be relaxed to allow for a $U(1)$ current algebra, and we expect it is ultimately necessary to do so. However this current algebra is not the object of our present investigations (or those of \cite{Compere:2009dp}) so for simplicity we will continue to suppress it with these slightly stronger boundary conditions.}\be \d A_t \sim \O(r) \;,\;\;\;\; \d A_r \sim \O(r^{-2})\;,\;\;\;\; \d A_\th \sim \O(1)
 \;,\;\;\;\; \d A_{\tilde{y}} \sim \O(r^{-1}) \;,\;\;\;\;\d A_{\phi} \sim \O(r^{-1})\ee
These boundary conditions lead to  an asymptotic symmetry group consisting of a Virasoro algebra 
generated by the following diffeomorphisms and gauge transformations
\be\label{kcv}
\xi(\e )= \e (\tilde y )\p_{\tilde y } - \p_{\tilde y }\e(\tilde y )\,r\p_r +\Theta \e(\tilde{y})f.
\ee
Here 
\be
\Theta \equiv  {a  e^\d \over 2\pi T_R} \, \tanh 2 \d 
\ee
is the background value of $- A_{\tilde{y}}$ in (\ref{aeq}).
The last term generates the gauge transformation $\delta A= -\Theta\, d\e(\tilde{y})$. 
We note that for constant $\e$ we recover  the gauge invariant rotation generator (\ref{jdir})
\be
 \xi(\pi T_R)= j_L.
 \ee
Given the identification (\ref{diu}),  the zero mode of $\xi$ is then twice the total angular momentum in the NHEK
region. Note that this is not the same as the angular momentum $J_L$ of the asymptotically flat black hole, as the latter
receives contributions from outside the NHEK region\footnote{ Indeed, for the supersymmetric BMPV case, all of the angular momentum is stored outside
the near-horizon region \cite{Gauntlett:1998fz}.}.
 
The Lie bracket algebra  is
\be
\{ \xi(\e), \xi({\e'})\}_{L.B.} =  \,\xi(\e\p_{\tilde y }\e'-\e'\p_{\tilde y }\e)
\ee
The Dirac bracket algebra of the associated generators is \be \label{vrs}
\{Q(\xi),Q(\xi')\}_{D.B.} =Q_{\{ \xi, \xi' \}_{L.B.}}+  \int k_{\xi} (\mathcal{L}_{\xi'} g, g)
\ee
where the expression for $k_{\xi}$ can be found in \cite{barnichbrandt,barnichcompere}.  This last term gives a central extension of the Virasoro algebra  with
\be \label{cr}
c_R = 6 J_L.
\ee
Using this value of $c_R$ and the temperature $T_R$ conjugate to angular momentum (\ref{tr}),  the microscopic Cardy formula then reproduces the macroscopic area law

\be S_{micro}={\pi^2c_R T_R \over 3}=2\pi\sqrt{J_L^2-Q^3}=S_{BH}.\ee

\section{CFT interpretation}

Now we wish to describe the holographic duals of the geometries (\ref{ff}).  We begin with the maximal case, where Kerr/CFT reduces to AdS/CFT and well-understood stringy methods can be applied. 
We will see that AdS/CFT methods can also be easily used
to understand  the linearized deformations, but more work will be required for the general case.

\subsection{The maximal case}
 For $\delta \to \infty$ with fixed $J_L$, $a \to Pe^{-\delta}$,  $Q^3\to J_L^2=P^6$, and we have a maximally charged
 5D Kerr-Newman black hole. This solution can also be reached by increasing the spin of the supersymmetric  BMPV black hole up 
to its maximal value, and was studied as such in \cite{mgas,gibb,ldys}. In the limit $\delta \to \infty$,
\be
\alpha \to 0 \ , \quad \gamma \to 1.
\ee
The last term in (\ref{nicemetric}) vanishes and we recover locally AdS$_3\times S^3$ as the total space of the $U(1)$
bundle over the $AdS_2\times S^3$ 5D black hole.

The generators of the enhanced  local $ SL(2,R)_R$ which appears at maximality are
\bea\label{rght}
\bar H_n &=& f_n(y)\p_y - r f_n'(y) \p_r - {1\over r}f_n''(y)\p_t  \non \\
 & & \quad f_n(y) = e^{-ny} \ , \quad n=0,\pm 1 
\eea
while the global $SL(2,R)_L $ generators are 
\bea\label{isom1}
 H_n &=& -(g_n(t) + {1\over 2 r^2}g_n''(t))\p_t  + r g_n'(t)\p_r + {1\over r}\, g_n''(t)\p_y \non \\
& & \quad g_n(t) = t^{n+1} \ , \quad n=0,\pm 1 
\eea
There is also a local enhanced $SU(2)_L$
\bea\label{isom2}
J^1 &=& \cos\psi\, \p_\theta + {\sin\psi\over \sin \theta}\,\p_\phi - \sin\psi \cot \theta \,\p_\psi\\
J^2 &=& - \sin \psi\, \p_\theta +  {\cos\psi \over \sin \theta}\,\p_\phi - \cos\psi \cot\theta \,\p_\psi\notag\\
J^3 &=& \p_\psi \ .\notag
\eea
The global $SU(2)_R$ generators are given by a similar expression with $\psi \leftrightarrow \phi$.

Tn the $\delta \to \infty$ limit, one finds
\be T_Q  \to {3 e^{-2\delta} \over  \pi P },~~T_R  \to  {3e^{-2\delta}\over  \pi},~~f \to -  { 2 \over P}\,\p_\psi, ~~
j_L \to 3\,e^{-2\delta} \,\p_y 
\ee
Note that  in this limit  the fiber vector $f$  is tangent to the $S^3$, while the angular momentum $j$ is tangent to the AdS$_3$ factor!
The identification (\ref{id}) becomes
\be y \sim y,~~~~~\psi \sim \psi -{4\pi m \over P} +4\pi n \ee
Since $P\in \mathbb{Z}$, in the limit taken this way  the $4\pi n$ identification is redundant. The approach to the limit however is subtle and will be discussed in the next subsection.

This geometry also arises as the near horizon of the $P^3$ string in 5D,
which is locally a $U(1)$ bundle over $AdS_3\times S^2$.  To see this we rewrite the metric at $\delta=\infty$ in the form
\be \label{fas}ds^2={Q \over 4} (-w^+w^- +w_3^2 +d\theta^2+\sin^2\theta d\phi^2)+(du+{\cal A})^2,\ee
where  ${\cal A}=-{P\over 2} \cos\theta d\phi$ and we have defined a canonically identified coordinate
\be u=-{P \over 2}\,\psi \sim u+2\pi m. \ee
This is manifestly a $U(1)$ bundle parameterized by $u$ over $AdS_3\times S^2$.
 Integrating the magnetic
graviphoton field strength ${\cal F}=d{\cal A}$ over the $S^2$ we see that this corresponds to a magnetic string solution with charge $P$.
The holographic dual of this geometry is of course the $P^3$ CFT, which therefore must also describe maximal extreme Kerr-Newman.

In $AdS_3$, the $SL(2,R)_R$ isometry algebra (\ref{rght}) is enhanced to a Virasoro which in the geometry above has $c_R=3\ell /2G_3=6P^3$.
With the boundary conditions of \cite{pw}  these are generated by the vector fields 
\be
  \xi_{AdS}(f)=f(y)\p_y-\p_yf(y)r\p_r+{\cal O}(r^{-1}) \p_t +{\cal O}(r^{-1}) \p_y +{\cal O}(1) \p_r  ,
   \ee
where the correction terms are trivial.\footnote{The boundary conditions of \cite{pw} are more relaxed than the original ones of \cite{Brown}, 
but nevertheless consistently lead to the same ASG and central charge. Note that the $r$ coordinate here is not the same as the one of \cite{pw}.  } 
 One expects this Virasoro to be directly related to some  limit of the Kerr-CFT Virasoro. Indeed,  transforming the generators of the latter, given in (\ref{kcv}),  to $(y,\psi)$ coordinates
\bea
 \xi(\e)&=& \e (\tilde y  )(\p_{\tilde y }  + \Theta f) -  \p_{\tilde y }\e(\tilde y )r\p_r \\
&=& \e ( {y +\pi T_R \psi} )\left( \frac{1- 2 \pi T_Q \Theta}{\pi T_R}\p_{\psi}+ 
2 \pi T_Q  \Theta \, \p_y\right) -  \p_y\e({y + \pi T_R \psi}   )r\p_r ,
\eea
and taking
\be  \e(x   )=f(x), \ee
one finds
\be \lim_{T_R\to 0} \xi(\e)=\xi_{AdS}(f).\ee 
Hence we conclude that the standard right-moving Virasoro of $AdS_3$ coincides in the maximal limit with the Kerr-CFT Virasoro.

\subsection{Explicit brane construction}

The discussions of the previous sections pertain to essentially any M/string compactification with a consistent truncation to 5D minimal supergravity.  Although the dual CFT will always have central charge $6P^3$ (to leading order), the details of the CFT will depend on the compactification.  In this section we give an explicit example of a brane construction. 

The 6D metric (\ref{fas})  can be uplifted to a solution of type IIB string theory by interpreting (\ref{hf}) as the RR three form field strength and  adding a $T^4$ of unit radius with coordinates $(x^6,x^7,x^8,x^9)$ to the geometry. This is the near horizon geometry of a certain brane configuration in the compactification to 5D on $(u, x^6,x^7,x^8,x^9)$ (i.e. $T^4$ and the Hopf fiber of the $S^3/Z_P$). The numbers of branes of each type can be determined by computing the charges. The number of D5 branes wrapping the $T^4$ is
\be Q_5= \frac{1}{4 \pi^2} \int_{S^3}  H =P. \ee
The number of D1 branes dissolved in the D5 branes is 
\be  Q_1= \frac{1}{(2 \pi)^6} \int_{S^3 \times T^4 } \star H =  P .\ee
In addition, the fact that the $S^1$ parametrized by $u=-\half P \psi $ is fibered over the $S^2$
implies there is a charge $Q_{TN}=P$ magnetic Taub-NUT string
\be
Q_{TN} = \frac{1}{2\pi} \int_{S^2} {\cal F} = P 
\ee
Therefore, the brane configuration that produces the geometry at maximality consists of $P$ copies of 

\medskip

\begin{center}
\begin{tabular}{c|cccccc} 
  & 6 & 7& 8 & 9 & $u$ & $ \mathbb{R}_{y} $  \\ \hline
D5 & X & X& X & X &  & X  \\
D1 & & & & & & X  \\
TN & & & & &  &X 
\end{tabular}
\end{center}
\medskip
where $\mathbb{R}_{y} $ denotes the common string direction. 
This can be dualized to three intersecting stacks of 
M5 branes in M-theory as follows: first, T-dualize on  and $x^8$,$x^9$ and $u$, which yields  the following configuration
in type IIA

\medskip

\begin{center}
\begin{tabular}{c|cccccc} 
  & 6 & 7 & 8 & 9 & $u$ & $\mathbb{R}_{y} $  \\ \hline
D4 & X & X&  &  & X  & X  \\
D4 & & & X & X & X & X  \\
NS5 &X &X &X & X&  & X  
\end{tabular}
\end{center}
\medskip
Lifting to  M-theory with an extra coordinate $x^{10}$, we obtain three intersecting  stacks of $P$ M5 branes with a common noncompact direction $y$: 

\medskip

\begin{center}
\begin{tabular}{c|ccccccc} 
  & 6 & 7 & 8 & 9 &10& $u $ & $\mathbb{R}_{y} $  \\ \hline
M5 & X & X&  &  & X &X & X  \\
M5 & & & X & X & X & X& X  \\
M5 &X &X &X & X&  & &  X  
\end{tabular}
\end{center}
\medskip
This gives an MSW string for a $T^6$ compactification of $M$ theory. The self intersection number of this stack of M5 branes is $6P^3$, giving the central charges \cite{msw}
\be
c_L = c_R = 6 P^3.
\ee 
\subsection{Linear deformations}

As mentioned above, there are some subtleties in the $\delta\to \infty$ limit. These can be understood by doing
an expansion in $\alpha \sim 4 e^{-2\delta}$ around the maximal solution.  Since $\gamma
\sim 1+{\cal O}(e^{-4\delta})$, at this order the only deformation of the geometry  (\ref{fas}) is in the gauge field

\be \label{resx} { \cal A} \r  {\cal A } -{P \over 2}\, \alpha  w_3, \ee
which leads to a radial  electric field ${\cal F}_{el}=2 Pe^{-2\delta} dr \wedge dt$ in addition to the magnetic
one. This deformation
should have a dual interpretation in the $P^3$ CFT, presumably as a source for some CFT operator or deformed state. Since it is a deformation of $AdS_3$ at this order, this can be analyzed with standard
methods and will be discussed in \cite{ghss}.

In addition, there is a deformation
of the identification (\ref{id}).  At large $\delta$
 \be \label{xxd} y \sim y+{12 \pi m e^{-2\delta}\over P},~~~~~\psi \sim \psi-{4\pi  m\over P}+4\pi n, \ee
To clarify the structure of the lattice of identifications generated by $(m,n)$ it is convenient to redefine
$m=\hat m + nP$. To leading order in $e^{-2\delta}$, (\ref{xxd}) then becomes \be \label{xxc} y \sim y+{12 \pi n e^{-2\delta}},~~~~~\psi \sim \psi-{4\pi  \hat m\over P}, \ee
The $\psi$ identification was already discussed in the previous subsection and produces the geometry of a charge $P$ magnetic string. What is the meaning of this $y$ identification? In general the quotient of $AdS_3$
 by the $SL(2,R)_R$ element
 \be e^{4\pi^2  i T_R \bar H_0 }   \ee
 is dual to a thermal state in the CFT at dimensionless right-moving temperature $T_R$.  From (\ref{rght}) we see that shifts in $y$ are generated by $\bar H_0$, and the CFT is therefore at temperature
\be T_R\sim  {3 \,e^{-2\delta}\over \pi } ,\ee
This is just the linearization in $e^{-2\delta}$ of
the temperature derived in (\ref{tr}) as the thermodynamic conjugate to angular momentum.
According to  Cardy's formula, the entropy to linear order is
\be\label{ddx} S_{micro}={\pi^2 c_RT_R \over 3}\sim 6\pi J_L e^{-2\delta}.\ee
This agrees exactly with the linearization of the area law (\ref{sbzh}).

Note that, since the temperature itself is already linear in the deformation, the entropy computation (\ref{ddx}) does not require any knowledge of the deformation of the CFT induced by the operator dual to (\ref{resx}). This will however be required at quadratic and higher order.

\subsection{Finite deformations}

At second order, the electric field strength backreacts on the geometry. This breaks the $SL(2,R)_R$  and the $AdS_3$ geometry becomes  warped.
The finitely deformed geometry (\ref{nicemetric})  is a solution of M/string theory and can be reached
by a continuous deformation of the $AdS_3$ dual of the $P^3$ CFT. Therefore it should have a holographic dual which can itself be reached by a continuous deformation of the $P^3$ CFT. The analysis of the asymptotic symmetry group suggests that conformal invariance persists in some form for finite deformations, and that the central charge is undeformed to leading order at large $J_L$. It is therefore natural to conjecture that the dual of the finitely deformed geometry \eqref{nicemetric} is simply a finite deformation of the $P^3$ CFT. This deformation will involve both the dynamics and the (mixed) state itself.
We note that the finitely deformed geometry \eqref{nicemetric} cannot be glued on to an asymptotically flat 5D string region\footnote{It is possible that this geometry can be obtained in a near horizon scaling limit of the general black string geometry in which the asymptotically measured M2 charge goes to infinity.} (although of course it can be glued on to an asymptotically flat 5D Kerr-Newman). Since everything has been embedded in string theory, this type of warped deformation should have a  sensible holographic dual, but it has not been considered in the literature. Closely related deformations were considered  in
 \cite{Duff:1998cr,Israel:2003cx,Israel:2004vv,Detournay:2005fz,Anninos:2008qb}, see especially the recent paper 
\cite{Detournay:2010rh} in which a worldsheet construction is presented. Clearly, a good understanding of this novel type of deformation from the CFT perspective is highly desirable. We hope to return to this elsewhere \cite{ghss}.

\section*{Acknowledgements}
We are grateful to Miranda Cheng, Geoffrey Comp\`{e}re, M. Gaberdiel, Tom Hartman, Josh Lapan, 
Juan Maldacena, Wei Song, C. Vafa  and Xi Yin
for useful conversations. 
We would like to thank the Institut d'\'Etudes Scientifiques de Carg\`{e}se  and the Centro Stefano Franscini in Ascona, Monte Verit\`{a}, where part of this work was carried out, for providing a productive research environment. 
 This work was supported in part by the Fundamental Laws Initiative at
 Harvard and DOE grant DE-FGO29ER40654 and by ANR grant 08-JCJC-0001-0, and 
the ERC Starting Independent Researcher Grant 240210 - String-QCD-BH.

\appendix

\section{Long vs. short string pictures}

For a given black hole, one expects that many different Virasoro algebras exist, with differing central charges 
associated with different choices of circle in the geometry. This arises already
in the study of superysmmetric black holes in toroidal compactifications
of string theory. For example, in the case of the 5D black hole with $Q_5$ NS5-branes, $Q_1$ 
fundamental strings wrapping an $S^1$ and carrying momentum $n$, T-duality changes the central 
charge from $6Q_1Q_5$ to $6nQ_5$.  Of course the temperature also changes and the entropy is always
 a U-duality invariant - see e.g. \cite{Duff:1998cr} . Similarly, in discussions of Kerr-CFT, 
when there are multiple angular momenta there are multiple sets of consistent
 boundary conditions with different central charges and temperatures but the same entropy 
\cite{Azeyanagi:2008kb,Chow:2008dp}.
 When a $U(1)$ charge is present and the fiber is a geometric circle, there are still more 
possibilities \cite{Hartman:2008pb}. 

In the present context, there  is an alternate and natural computation  of the Kerr-Newman entropy using the
 Virasoro symmetry associated to
charge circle paramterized by $y$  at fixed $\psi$ in \eqref{nicemetric}.   The corresponding asymptotic symmetry 
group analysis then gives \cite{Azeyanagi:2008dk,Isono:2008kx}
\be 
\label{tas} c^Q = 6Q^2 .
\ee 
From the identification  (\ref{id}) the dimensionless temperature is $T_Q$. The Cardy formula then gives

\be 
S_{BH}={2\pi^2 }Q^2 T_Q=2\pi\sqrt{J_L^2-Q^3}.
\ee
Let us now try to understand the relation between this $Q$-CFT picture and the Kerr-CFT picture at least near 
the maximal point. The $Q$-CFT lives on the closed circle parameterized by
\be 
y= 2\pi T_Qs,~~~~0\le s\le 2\pi .
\ee
The fact that the identification (\ref{id}) also involves a shift of $\psi$ means that there is a finite chemical
 potential for the charge conjugate to $\psi$, or equivalently $\psi$-twisted boundary conditions. This does not
 affect the Cardy entropy. On the other hand, the Kerr-CFT lives on the circle
 \be 
\tilde y= 2\pi T_R s,~~~~0\le s\le 2\pi.
\ee
The ratio of the two $s$-coefficients is
\be 
{T_R \over T_Q}={Q^2\over J_L} \to P ~~{\rm for}~\delta\to\infty. 
\ee
Thus near maximality the Kerr-circle wraps $P$
times around the $Q$-circle and hence is  a "long-string" - in the sense of \cite{maldasuss} - of the $Q$-CFT.
The corresponding rescaled Virasoro generators are related by
\be L_{m}^{long}={ P}L_{m/P}, \ee
for $m$ a multiple of $P$.
It follows that
\be [L_m^{long},L_n^{long}]={P^2}[L_{m/P},L_{n/P}]=(m-n)L_{m+n}^{long}+{c \over 12 P}(m^3-ma)\delta_{m+n},\ee
where $a$ depends on the definition of the constant $L_0$ shift. Hence 
\be c^{long}= {c \over P}.\ee
This is precisely the relation between the Kerr/CFT and $Q$-CFT central charge at maximality. Note that moreover the
 temperature is rescaled by a factor of $P$ because the long string circle is $P$ times bigger. Hence, exactly as in
 \cite{maldasuss}, the product $c \,T$ appearing in the Cardy entropy is the same in long and short string pictures. 
Similar conclusions might be reached, pending a better understanding of the deformation,  more generally whenever 
the ratio ${T_R \over T_Q} = {Q^2 \over J_L}$ is rational.

 \end{document}